\documentclass[fleqn,usenatbib]{mnras}
\usepackage[utf8]{inputenc}
\usepackage{amsmath}
\usepackage{graphicx}
\usepackage{epstopdf}
\usepackage{threeparttable}
\usepackage{graphicx, color}
\usepackage{textcomp}
\usepackage{bm}
\usepackage{soul}
\usepackage{amssymb}
\usepackage{hyperref}
\usepackage{blindtext}

\newcommand{\eqb}{\begin{eqnarray}}
\newcommand{\eqe}{\end{eqnarray}}
\newcommand{\ergs}{erg~s$^{-1}$}

\newcommand{\fluxhz}{ergs cm$^{-2}$ s$^{-1}$ Hz$^{-1}$}

\title[Environment and properties of high redshift AGNs]
{Active Galaxy Nuclei at high redshifts: properties and environment of Type 1 and 2 AGNs}
\author[Bornancini et al.]{C. Bornancini$^{1,2}$\thanks{bornancini@oac.unc.edu.ar}, D. Garc\'ia Lambas$^{1,2}$\\
$^1$Instituto de Astronom\'{\i}a Te\'orica y Experimental, (IATE, CONICET-UNC), C\'ordoba, Argentina \\
$^2$Observatorio Astron\'omico, Universidad Nacional de C\'ordoba, Laprida 854, X5000BGR, 
C\'ordoba, Argentina\\
}

\begin{document}

\date{Received.../Accepted...}

\pagerange{\pageref{firstpage}--\pageref{lastpage}} \pubyear{2017}

\maketitle

\begin{abstract}
 
We explore host galaxy properties and environment of a sample of Type 1 and 2 active galactic nuclei (AGN) taken from the COSMOS2015 catalog, within 0.3 $\leq z \leq$ 1.1 selected for their emission in X-rays, optical spectra and SED signatures. 
We find different properties of host galaxies of distinct AGNs: Type 1 AGNs reside in blue, star-forming and less massive host galaxies compared to Type 2.
The majority of the AGNs have intermediate X-ray luminosities, $10^{42}\leq L_X<10^{44}$ \ergs, while only a few have X-ray luminosities  ($L_X > 10^{44}$ \ergs) as those observed in QSOs. 
Non-parametric morphological analysis show that the majority of Type 1 AGN hosts are elliptical or compact galaxies, while Type 2 AGN host present more scatter, from spirals, irregulars and elliptical galaxies. 
The environment of the different AGN types are similar except at small scales ($r_p<$100 kpc), where Type 2 AGNs have more neighbour galaxies than Type 1s.
Galaxies located close to Type 2 AGNs ($\sim$100 kpc) tend to have redder colours, and are more massive compared to the local environment of Type 1s.
The observed differences in the environment and host galaxy properties of Type 1 and 2 AGN types show that the obscuration due to the presence of gas and dust may be distributed in larger galactic-scales, possibly originated  by galaxy interactions or mergers.

\end{abstract}


\section{Introduction}

In recent years, the study of Active Galaxy Nuclei (AGN) have become a fundamental part of our understanding of the formation and evolution of galaxies and their systems, such as groups or cluster of galaxies \citep{miley,hatch,danner,cooke}.
It is clear now that most, if not all, galaxies with a bulge/spheroidal component harbor a super-massive black hole at their cores \cite[and references therein]{kor11}, even low-mass galaxies can harbor supermassive black holes \citep{filip,ahn}, therefore AGNs influence the formation and evolution of galaxies \citep{springela,springelb,hop}.
Several related issues such as the relationship between black hole mass and galaxy bulge mass \citep{mago,marconi03}, or velocity dispersion \citep{ferra,geb,tremaine,shen} and the relation between star formation history and the AGN activity through cosmic time \citep{dick,merloni04,kor13} provides further important clues on this topic.

\citet{anto} and \citet{urry} have proposed a unified scheme for the variety of AGNs.
In unified models, distinct spectral signatures between different AGN types would just the result of varying orientation relative to the line of sight. According to their optical spectra, certain AGN types are classified according to the presence of broad emission lines (FWHM $> 2000 \rm \, km \, s^{-1}$, Broad Line AGNs or Type 1) and narrow emission lines or by the absence of broad permitted emission lines (Narrow Line AGNs or Type 2).
In unified models it is expected that a large number of objects present a central torus-shaped obscured region (with sizes of a few parsecs) due to large amounts of gas and dust, that can block broad optical spectral lines produced in the near regions of the accretion disk. If the torus is face on, it is possible detect the broad-line region directly and the galaxies are identified as Type 1 AGN, otherwise, AGN observed through a torus will be classified as Type 2 AGN.
A question that be raised is whether the material that blocks the broad line region is confined to a small central region (few parsecs) in the form of a torus or if the obscuration can be produced by dust distributed at larger galactic scales or (both possibilities at the same time).

In the last years it has been proposed that evolutionary process could produce the obscuration observed in some AGNs.
In some models, obscured AGNs represent a phase in the co-evolution of the galaxy an its central black hole. Galaxy formation would be produced by mergers between gas-rich galaxies. In this first phase, the central nuclear region is completely obscured by large amounts of gas and dust. Then, there would be another phase where feedback process would sweep the obscuring material and the system enters in an unobscured, bright optical quasar phase \citep{hop,springela}.

Within the unified model paradigm, Type 1 and 2 AGNs only differ according to the line of sight with respect to the observer, no differences would be expected in host galaxy properties and they would be expected to have similar local galaxy environment. 
There are several works on analysis of the environment of AGNs, although the results obtained are contradictory.
\citet{jiang} studied the environments of Type 1 and Type 2 AGNs using samples of low-redshift (z $<$ 0.09) AGNs, normal galaxies and groups of galaxies selected from the Sloan Digital Sky Survey (SDSS). These authors found that Type 1 and Type 2 AGNs have similar clustering properties on large scales ($\geq$1 Mpc), but at scales smaller than 100 kpc, Type 2 have significantly more neighbours than Type 1 AGNs.
These results suggest that dark halos hosting Type 1 AGNs, on average, have similar masses to those hosting Type 2s, but that Type 1s have less number of surrounding satellites around them. Also, the distribution of Type 2 satellites is more centrally concentrated. These authors also find differences in the host galaxy colours. The optical colour distributions of the host galaxies of the two AGN types are similar, but their infrared colours are significantly different.


Differences between AGN environments are also found at high redshifts.
\citet{hickox11} studied a sample of mid-infrared obscured and unobscured quasars in the redshift range $0.7 < z < 1.8$, selected from the 9 deg$^2$ Bo\"{o}tes multiwavelength survey. These authors studied the spatial clustering of both AGN types and found a characteristic dark matter halo masses of log($M_{halo}$[h$^{-1}$ M$\odot$])$=12.7$ and 13.3 for unobscured and obscured AGNs, respectively.
Similar results were found by \citet{donoso} who calculated the angular correlation function for a sample of $\sim$ 170,000 AGNs extracted from the Wide-field Infrared Survey Explorer ({\it WISE}) catalog, selected to have a mid-IR colour cut (W1$-$W2 $>$ 0.8). They found that red (obscured) AGNs inhabit denser environments (with a typical dark matter halo mass of log($M_h$/$M_{\odot}$ $h^{-1}$) $\sim$13.5) than blue (unobscured) AGNs (with dark matter halo mass of log($M_h$/$M_{\odot}$ $h^{-1}$)$=12.8$).
In a similar way \citet{dipompeo} calculated the angular clustering of a sample of infrared-selected  obscured and unobscured quasars from the {\it WISE} survey, but applying a robust and conservative mask to WISE-selected AGNs. These authors found that obscured quasars reside in halos of higher mass: log($M_h$/$M_{\odot}$ $h^{-1}$)$\sim13.3$, while unobscured quasar host have log($M_h$/$M_{\odot}$ $h^{-1}$)$\sim12.8$. 
\citet{geach} present a cross-correlation analysis examining the relationship between the combined Type 1 and 2 quasar population at $z\sim 1$ and the mass using the cross-power spectrum of the projected mass density as traced by the convergence of the cosmic microwave background lensing field from the South Pole Telescope (SPT). These authors argue that the bias of the combined sample of Type 1 and 2 quasars is similar to that previously determined for Type 1 quasars alone. They conclude that obscured and unobscured quasars trace the mass in a similar way.
\citet{gilli} studied the spatial clustering of a variety of high redshift X-ray selected AGNs in the XMM-COSMOS field, adopting different definitions for the source obscuration. They do not find evidence that AGNs with broad optical lines (BLAGN) cluster differently from AGN without broad optical lines (non-BLAGN). Similar results are obtained when considering X-ray absorbed and X-ray unabsorbed AGNs.

In \citet{bornan17} we carried out a similar study to the one presented here, with a sample of mid-infrared selected AGNs, with spectroscopic and/or photometric redshifts, in addition to information in the X-rays. In this paper we presented a more intensive study, with a homogeneous and larger AGN sample obtained by combining X-ray, photometric and optical spectroscopic selections. We also study correlations with other parameters such as stellar masses and non-parametric measurements that allow estimates of host galaxy morphologies. In this paper we extend these studies and investigate the nature of Type 1 and 2 AGNs in order to study their host galaxies, its possible environmental dependence and their relation to unified models. 

This paper is organized as follows: In Section 2 we present datasets and samples selection, while in Section 3 we investigate about the properties of Type 1 and 2 AGN host galaxies. The environment of Type 1 and 2 AGNs and their neighbouring galaxies properties is analysed in Section 4. Finally, the summary and conclusions of our work are presented in Section 5. 
Throughout the work we will use the AB magnitude system \citep{oke} and we will assume the same $\Lambda$CDM cosmology adopted by \citet{laigle} with H$_{0} = 70$ km s$^{-1}$ Mpc$^{-1}$, $\Omega_{M} = 0.3$, $\Omega_{\Lambda} = 0.7$.

\section{Datasets}

Observational data analysed in this paper were obtained from the COSMOS2015 catalog \citep{laigle} which contains precise photometric redshifts and stellar masses
for more than half a million objects over the 2degre$^2$ COSMOS field\footnote{The catalog can be downloaded from ftp://ftp.iap.fr/pub/from\_users/hjmcc/COSMOS2015/} \citep{scoville}.
This catalog includes near-UV (0.23 $\mu$m) observations from GALEX \citep{zamo}, optical observations from the Canada-France Hawaii Telescope (u$^*$-band, CFHT/MegaCam; \citealt{sanders07}), and the COSMOS-20 survey, which is composed of 6 broad bands (B, V, g, r, i, z$^{++}$), 12 medium bands (IA427, IA464, IA484, IA505, IA527, IA574, IA624, IA679, IA709, IA738, IA767, and IA827), and two narrow bands (NB711, NB816), taken with Subaru Suprime-Cam \citep{tani07,tani15}.
Also near-infrared YJHKs-band data taken with WIRCam and Ultra-VISTA data \citep{mccrac10,mccrac12}. 
This field was also observed in the mid-infrared ([3.6], [4.5], [5.8] and [8.0] $\mu$m) from data obtained in the SPLASH-COSMOS survey (Spitzer Large Area Survey with Hyper-Suprime-Cam, PI: P. Capak), S-COSMOS \citep{sanders07}, the Spitzer Extended Mission Deep Survey and the Spitzer Candels survey data (Capak et al. 2015 in prep).

Several X-ray observations have been made in this field: XMM-COSMOS \citep{hasi07,cappe}, C-COSMOS \citep{elvis,civano16}, NuSTAR survey \citep{civano15}. The region has also been observed in other wavelengths, such as 24$\mu$m (MIPS, \citealt{lefloch09}), 160$\mu$m (PACS/PEP, \citealt{lutz}), 250, 350 and 500$\mu$m (SPIRE/HERMES, \citealt{oliver}), and public previous radio data at 20 and 90 cm. An exhaustive overview of the COSMOS field and multiwavelength data products is available at: http://cosmos.astro.caltech.edu/

\subsection{Type 1 and Type 2 AGN selection}
\label{agn}

\begin{figure}
 \centering 
\includegraphics[width=0.47\textwidth]{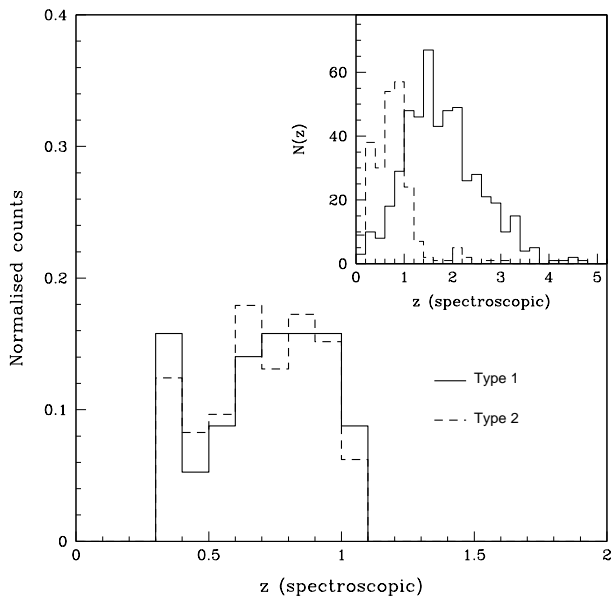}
	\caption{Normalised redshift distribution for the final selected Type 1 (solid line histogram) and Type 2 (dashed lines) AGNs. The inset shows the original redshift distribution of the mentioned AGN samples.}
\label{z}
\end{figure}

The AGN sample selection is based on the COSMOS2015 parent catalog. We also consider AGN spectral features and the presence of X-ray emission obtained from X-ray and spectroscopic data given by \citet{marchesi}. 
These authors identified 1770 X-ray sources from the master spectroscopic catalog available for the COSMOS collaboration (M. Salvato et al., in preparation).
They performed a spectroscopic classification of this X-ray sources sample as follows: sources that present at least one broad line (FWHM $> 2000 \rm \, km \, s^{-1}$) in their spectra  were classified as Broad Line AGN (BLAGN, 36\% of the spectroscopic sample) and sources with only narrow emission lines or absorption lines are defined as “non broad-line AGN” (non-BLAGN, 59\% of the spectroscopic sample).
Since the majority of these sources have low S/N spectra, or are in an observed wavelength range which does not allow to use emission line diagnostic diagrams, these authors do not make a further separation between star-forming galaxies and Type 2 AGN.
\citet{marchesi} also provide photometric identification from SED fitting technique for 3885 objects (96\% of the total X-ray sample). 
They used the method described by \citet{salvato11}, which adjusts templates to the sources multiwavelength SEDs.
The templates are divided in \citep{salvato09}: `unobscured AGN', which corresponds to a type I AGN or type I QSO template, `obscured AGN', which corresponds to a type II AGN or type II QSO template, `galaxy', which corresponds to a elliptical, spiral, or starburst galaxy, and `star'.
\citet{marchesi} found a general agreement between spectroscopic and photometric AGN classifications. These authors found that 82\% of the sources with BLAGN spectral type have been fitted with an unobscured AGN template, while 97\% of the non-BLAGN are fitted with either a galaxy template (74\%) or with an obscured AGN template (23\%).
In this paper we have selected a sample of AGNs according to the following definitions: Type 1 (unobscured) AGN from spectroscopic identification (1=BLAGN, column 26 in Marchesi's catalog\footnote{The catalog can be downloaded from http://irsa.ipac.caltech.edu/data/COSMOS/tables/chandra/}), Type 2 (obscured) AGN (2=non-BLAGN, column 26) and the photometric identification from SED (2$=$obscured, column 45).
Although the redshift distribution of both Type 1 and 2 samples are fairly different, we have selected a subsample of AGNs at z$\lesssim 1.1$ for which the distributions are comparable (see Figure \ref {z}). This selection yields samples of 145 Type 1 and 57 Type 2 AGNs in the redshift range 0.3 $\leq z \leq$ 1.1.




\section{Host galaxy properties}

\subsection{Mid-infrared colours}

It is well-known that AGN are preferentially located in some particular regions of the colour-magnitude diagrams \citep{hickox11,hickox07,bornan17}.
\citet{hickox07,hickox11} showed that IR-selected quasars show a bimodal distribution in optical to mid-IR colour ($R-[4.5]$).
These authors use this feature to define two different populations, one consisting of obscured objects
in the optical and the other of AGNs with similar properties to Type I QSOs (unobscured). Similar results were obtained by \citet{bornan17} in a sample of obscured and unobscured AGNs by means of a simple optical-MIR colour cut criterion ($R -[4.5] = 3.05$.) with redshifts in the range 1 $\leq$ z $\leq$ 2.

\begin{figure}
 \centering 
\includegraphics[width=0.47\textwidth]{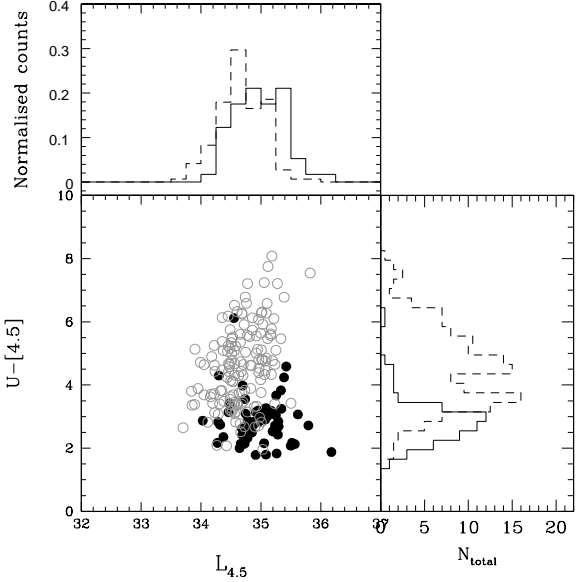}
\caption{Optical-IR colour, $U -[4.5]$ vs. L$_{4.5 \mu m}$ ($[4.5]$ microns luminosity for different AGNs with $0.3\leq z\leq 1.1$. 
Gray crosses and open circles represent Type 2 and Type 1 AGNs, respectively.	
	Left panel shows the corresponding colour distribution of different AGN types. Dashed line histogram represents the colour distribution of Type 2 AGNs, while solid line histogram correspond to Type 1 AGNs. Top panel shows the corresponding $[4.5]$ microns luminosity distribution of different AGN types (solid: Type 1, dashed: Type 2)}
\label{fig2}
\end{figure}

In Figure \ref{fig2} we show the $U-[4.5]$ colour vs. $[4.5]$ microns luminosity (L$_{4.5 \mu m}$) for the sample of AGNs with 0.3 $<z<$1.1 with X-ray emission and selected according to their spectral and photometric features.
The luminosity in 4.5 $\mu$m of each AGN, L$_{4.5 \mu m}$ was calculated by the following equation,

\begin{equation}
L_\nu(\nu_{\rm rest})=\frac{4\pi d_{\rm L}^2}{1+z} S_\nu(\nu_{\rm obs}),
\end{equation}

where $d_{\rm L}$ is the luminosity distance for a given redshift, $S_\nu$ is the flux density in \fluxhz,
and $\nu_{\rm obs}$ and $\nu_{\rm rest}$ are the observed and rest-frame frequencies respectively, where $\nu_{\rm
rest}=(1+z)\nu_{\rm obs}$. 

As it can be seen, Type 1 and 2 AGNs present different colour distributions. Type 1 AGN have blue $U-[4.5]$ colours in comparison with the Type 2 sample. Moreover, it is observed a bimodality in the colour distribution of Type 2 objects.

\begin{figure}
 \centering 
\includegraphics[width=0.47\textwidth]{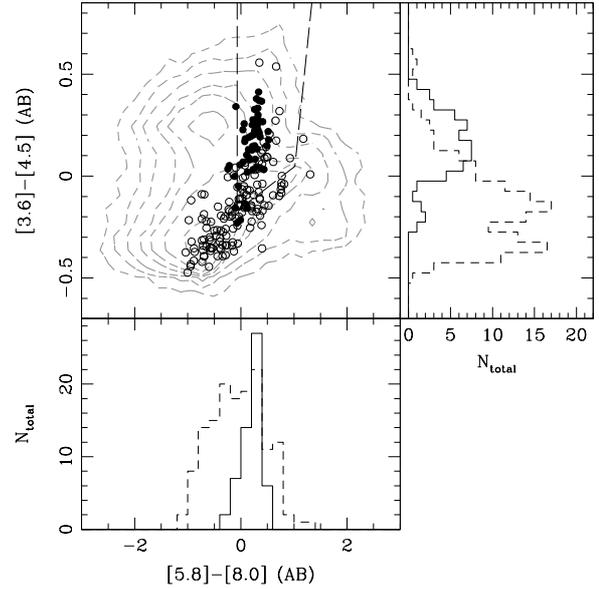}
\caption{Mid-infrared colour-colour diagram for sources with $\ge$3$\sigma$ detection in all four IRAC bands (contours). Open and filled circles represent the Type 2 and Type 1 AGN sample, respectively. The wedge marked with dashed lines correspond to the AGN selection criterion of \citet{stern05}}
\label{fig3}
\end{figure}

In the last years, several methods have emerged in order to identify AGNs from their colours at IR wavelengths.
One of these methods was proposed by \citet{stern05} who used mid-infrared observations with the Infrared Array Camera (IRAC) mounted on the Spitzer Space Telescope. 
In Figure \ref{fig3} we show the $[3.6]-[4.5]$ versus $[5.8]-[8.0]$ colour-colour magnitude diagram for sources brighter than 3$\sigma$ detection limits\footnote{3$\sigma$ depth in $m_ {AB}$ computed from the RMS maps, after masking the area containing an objects based on the segmentation map.} in all IRAC four bands ($[3.6]=25.5$, $[4.5]=25.5$, $[5.8]=23.0$ and $[8.0]=22.50$, \citealt{sanders07}). We also plot the corresponding colours of Type 1 (black filled circles) and 2 (open circles) AGNs. As it can be seen, the colour distribution are significantly different. Type 1 AGNs are preferentially located within the Stern et al. wedge (dashed lines), although a large fraction of Type 2 objects are located outside the AGN selection boundaries. For Type 1 AGNs we find that 86\% of the total sample are located inside the Stern et al. wedge. On the other hand, we find that only 28\% of Type 2 AGNs are located within this limits.
We also plot in this figure the colour distribution of each AGN type. In the right panel, we show the $[3.6]-[4.5]$  colour distribution of Type 1 (solid line histogram) and Type 2 (in dashed lines). In the case of Type 2 objects, we find a bimodality in the colour distribution, in a similar way of that found in the $U-[4.5]$ colour (see Figure \ref{fig2}), but mid-IR colours appear inverted, i.e. Type 1 AGNs are bluer comparing to the Type 2 sample. In the lower panel of Figure \ref{fig3} we plot the  $[5.8]-[8.0]$ colour distribution. As it can be seen both distributions are similar, except that Type 2 AGN colours are, on average, redder than those of Type 1 AGNs.


In order to quantify the presence of a colour bimodality, we perform a Gaussian mixture modeling (GMM) test.
We have used the GMM code of \citet{muratov} to quantify the probability that the colour distributions are better described by a bimodal rather than a unimodal distribution.
This code uses information from three different statistic tools: the kurtosis, the distance from the mean peaks ($D$), and the likelihood ratio test (LRT).
The kurtosis test measures the degree of peakedness of a distribution: a positive value corresponds to a sharply
peaked distribution whereas a negative kurtosis corresponds to a flattened distribution.
The distance from the mean peaks or ``Bandwidth test'', is the separation between the means of the Gaussian components relative to their widths, calculated as
\begin{equation}
D=\frac{\big|\mu_1-\mu_2 \big|}{\sqrt{\frac{\sigma^2_1+\sigma^2_2}{2}}}
,\end{equation}

where $\mu_1$ and $\mu_2$ are the mean values of the two peaks of the proposed bimodal distribution, and $\sigma_1$, $\sigma_2$ are the corresponding standard deviations. For a clear separation between the two peaks it is required that $D > 2$.

LRT is defined as

\begin{equation}
LRT=2\times\left[ln(L_{bimodal})-ln(L_{unimodal})\right]
,\end{equation}

where $L_{bimodal}$ and $L_{unimodal}$ are the likelihood for a bimodal and unimodal distributions, respectively.

\begin{figure}
\centering 
\includegraphics[width=0.47\textwidth]{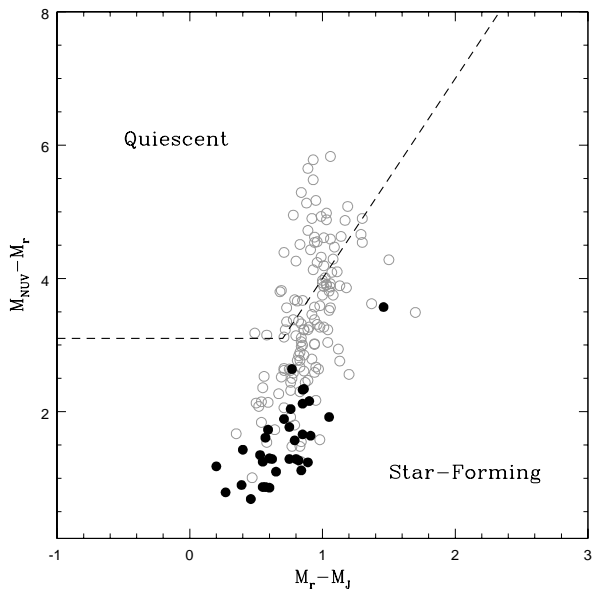}
\caption{Rest-frame $M_{NUV}-M_R$ vs. $M_R-M_J$ colours of Type 1 (filled circles) and Type 2 (open circles) AGNs. Dashed lines show the limits used to empirically separate quiescent from star-forming galaxies taken from \citet{ilbert13}.}
\label{quies}
\end{figure}

The GMM code provides the probability $p(\chi^2)$ of rejection of a unimodal distribution in favour of a bimodal fit using initial values for the two estimated peaks in the observed distribution.
In order to reach a maximum value of $p(\chi^2)$ we have performed tests with different values of $\mu_1$ and $\mu_2$.

For the $U-[4.5]$ colour distribution of Type 2 AGNs shown in the right panel of Figure \ref{fig2} (shaded line histogram), we find $\mu_1=3.5\pm0.4$ and $\mu_2=5.3\pm0.7$, with $p(\chi^2)=0.05$, $D=2.1\pm0.5$, kurtosis$=-0.39$. The value of $p(\chi^2)$ indicates that there is 5\% probability of the colour data being drawn from a single Gaussian model, rather than the best-fit double Gaussian model.

For the $[3.6]-[4.5]$ colour distribution of Type 2 AGNs shown in the right panel of Figure \ref{fig3}, we find $\mu_1=-0.08\pm0.05$ and $\mu_2=-0.34\pm0.04$, with $p(\chi^2)=0.004$, $D=1.9\pm0.2$, kurtosis$=0.97$. We notice however, that although $p(\chi^2)$ is very close to 0, indicating bimodality, the kurtosis is positive.
This analysis shows that the colour distribution of Type 2  AGNs is bimodal, and that these objects actually represent two different populations, one consistent with a blue host galaxy population and other consistent with a red host galaxy population. 



\subsection{Star-forming and quiescent galaxies}

The colours of galaxies in the near-UV ($NUV$), optical ($r-$band) and near-infrared ($J-$band) can be used to separate star-forming from quiescent galaxies. \citet{williams} investigated the properties of galaxies with $K<22.4$ (AB) and $z_{phot}\lesssim 2.5$ selected from the UKIDSS Ultra-Deep Survey, Subaru-XMM Deep Survey and Spitzer Wide-Area Infrared Extragalactic Survey. These authors found two distinct population of galaxies in the rest-frame $U-V$ vs. $V-J$ colour space: a clump of red, quiescent galaxies and a track of star-forming galaxies. From an angular cross-correlation analysis, \citet{williams} found that quiescent galaxies are clustered more strongly than those actively forming stars, indicating that galaxies with early-quenched star formation may occupy more massive host dark matter halos. Quiescent galaxies are possibly the progenitors of the most massive, early-type galaxies found locally, although evolutionary mechanisms are poorly understood.

\citet{ilbert10} used a slightly modified version of the colour-colour selection technique proposed by \citet{williams}. They used the colour $NUV-r$ instead of $U-V$ since this colour is a better indicator of the star formation activity.

In Figure \ref{quies} we plot the rest-frame colour-colour $NUV-r$ vs. $r-J$ for the AGN sample analysed in this work.
Quiescent objects are those with $M_{NUV}-M_r > 3(M_r-M_J)+1$ and $M_{NUV}-M_r > 3.1$ (dashed lines) while star-forming galaxies occupy regions outside of this criterion\citep{ilbert10, ilbert13}.
As noted by \citet{ilbert13} this technique avoids mixing the red dusty galaxies and quiescent galaxies. 
We find that Type 1 objects are located in the region populated by star-forming galaxies, whereas Type 2 AGNs are divided between the region occupied by quiescent (29\%) and star-forming galaxies (71\%).

\begin{figure}
 \centering 
\includegraphics[width=0.47\textwidth]{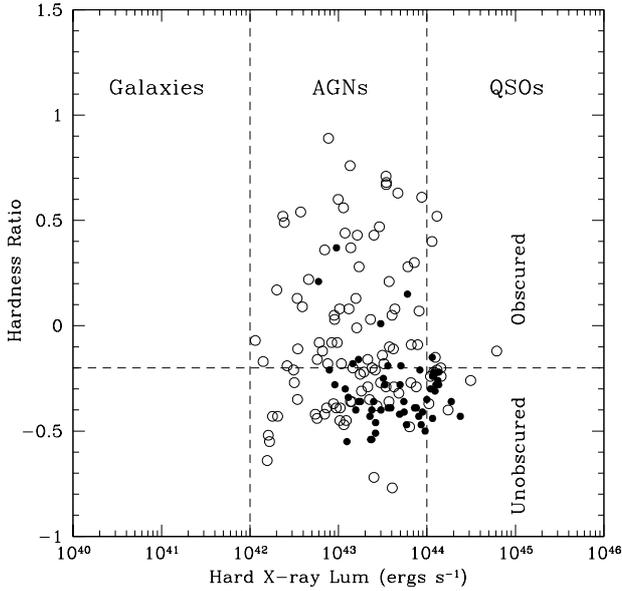}
\caption{Hardness ratio as a function of hard X-ray luminosity for Type 2 (open circles) and Type 1 (filled circles) AGNs. 
Vertical dashed lines show the typical separation for normal galaxies, AGNs and quasars used in the X-rays. The dashed horizontal line shows the HR value for a source with $N_H>10^{21.6}$ cm$^2$ at $z>1$. }
\label{HR}
\end{figure}

\subsection{X-ray properties}

\begin{figure*}
\centering 
\includegraphics[width=0.47\textwidth]{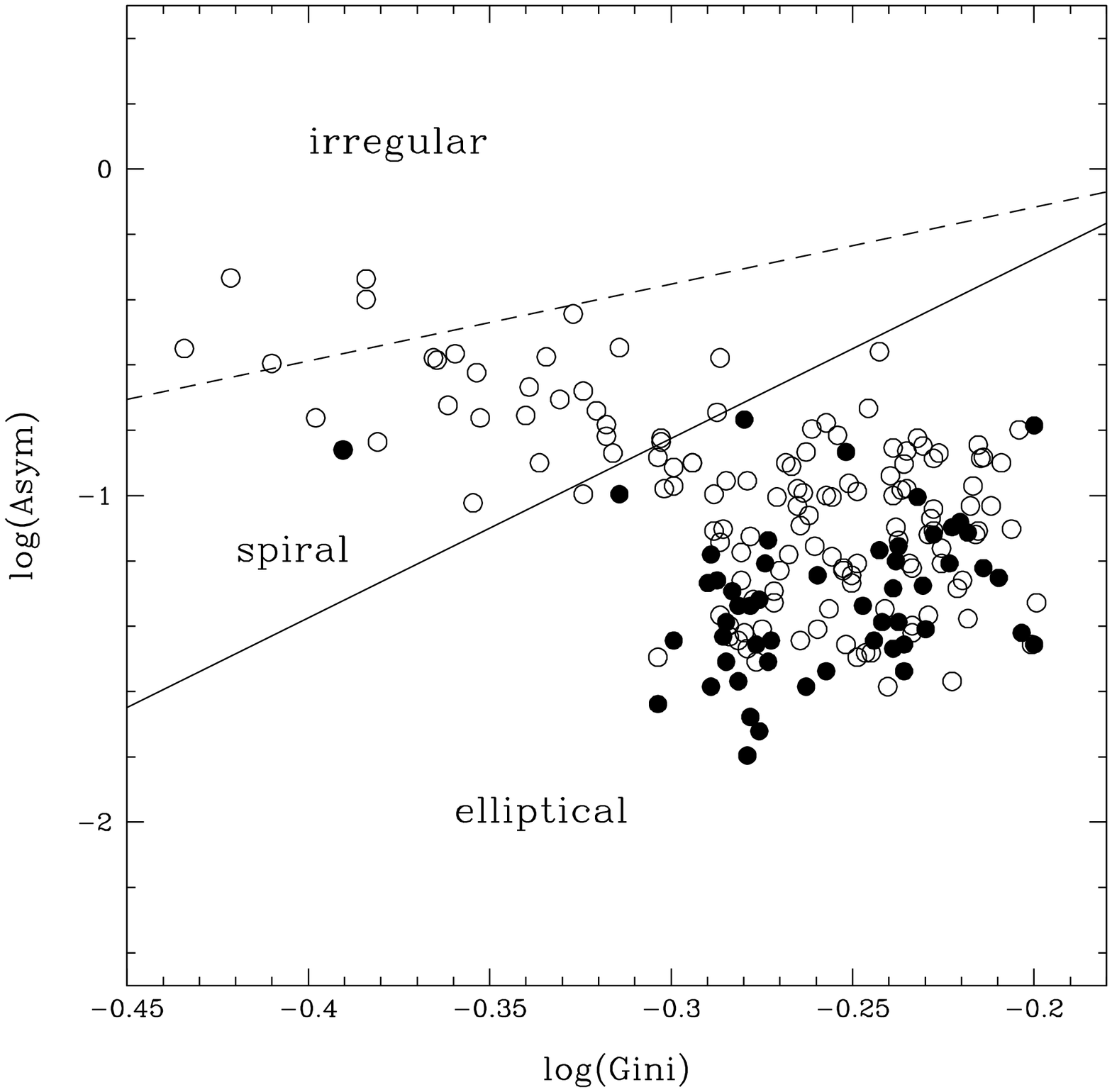}
\includegraphics[width=0.47\textwidth]{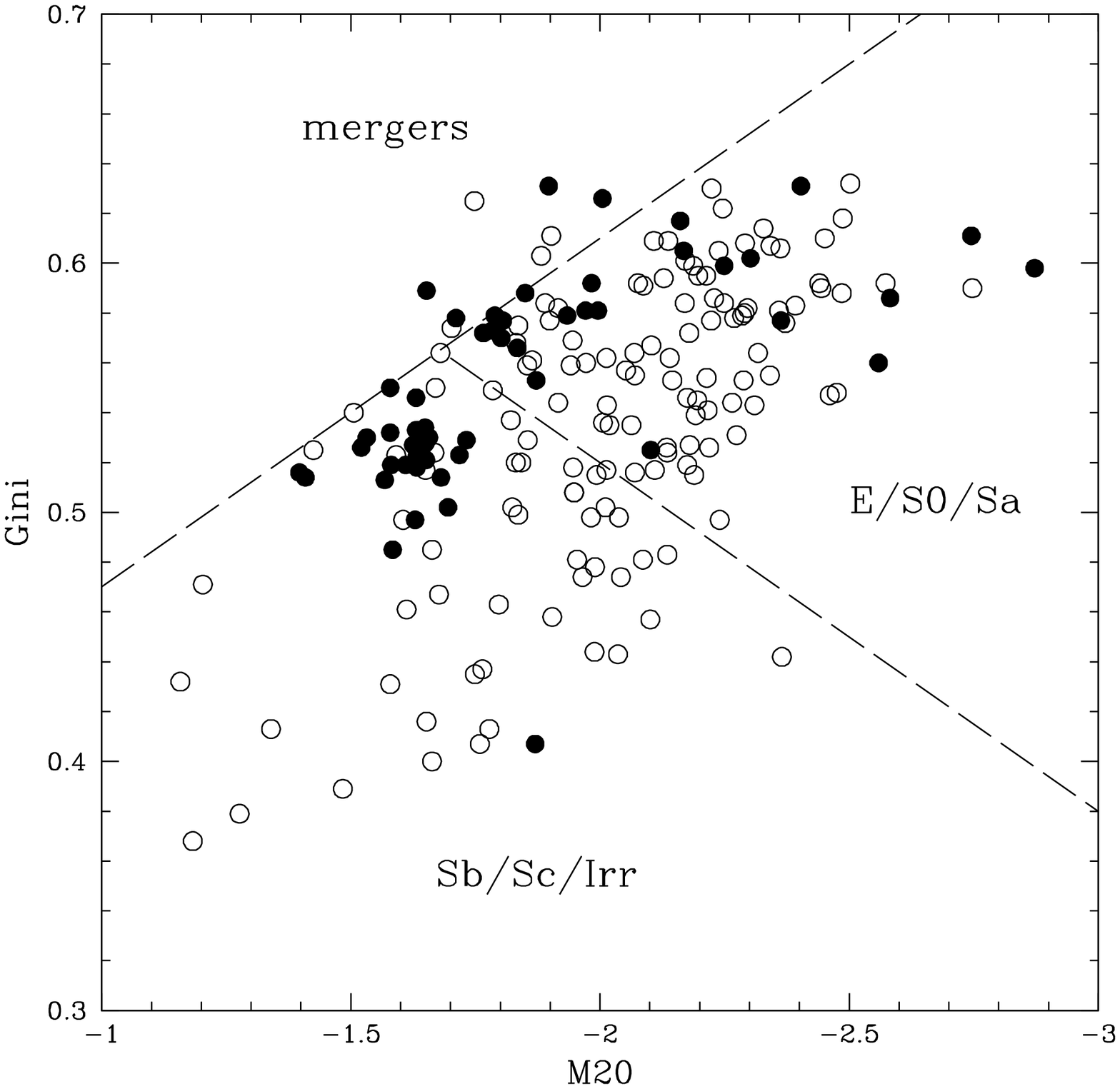}
	\caption{Left panel: Gini coefficient (G) vs. rotational asymmetry (A) for Type 1 (filled circles) and Type 2 (open circles). Dashed lines shows regions of predominantly irregular, spiral, and elliptical types from \citet{capak}. Right panel: M20 vs. Gini coefficient (G). The symbols are the same as in the left panel. Dividing lines show regions of mergers, Sb/Sc/Irr and E/S0/Sa galaxy types taken from \citet{lotz08}.}
\label{giniasym}
\end{figure*}

In this section we explore the X-ray properties of Type 1 and 2 AGNs, relevant at understanding the population of obscured AGNs which harbor supermassive black holes actively accreting matter. 
Figure \ref{HR} shows the hardness ratio (HR) as a function of hard X-ray luminosity ($2-10$ keV) for the sample of Type 1 and 2 objects selected in this paper.
The hardness ratio (which is an indication of the X-ray spectral shape) is defined as follows,

\begin{equation}
{\rm HR}=\frac{H-S}{H+S},
\end{equation}
where $H$ and $S$ are the count rates in the hard ($2-10$ keV) and soft ($0.5-2$ keV) bands, respectively. 

The dashed horizontal line shows the HR value (HR$=-0.2$) for a source with a neutral hydrogen column density, $N_H>10^{21.6}$ cm$^2$ at $z>1$ \citep{gilli}, which is used by several authors \citep{gilli, treis09, marchesi} to separate obscured and unobscured sources in the X-rays. This is due to the the fact that soft X-ray emission of obscured AGNs tend to be absorbed, while hard X-ray are able to escape.
As it can be seen, sources with hard X-ray spectra (positive hardness ratio) tend to be also classified as Type 2 sources in the optical, while Type 1 AGNs have in general a soft X-ray spectrum (negative hardness ratio). We find that most of the AGNs have X-ray luminosities similar to those found in  Seyfert-like galaxies (with $10^{42}\leq L_X<10^{44}$ \ergs), while only a minority have X-ray luminosities as the observed in QSOs ($L_X > 10^{44}$ \ergs). It can also be seen that there are no objects with $L_X < 10^{42}$ \ergs, which are usually identified with normal galaxies, in agreement with the AGN selection criteria proposed in Section \ref{agn}.

\subsection{Morphological analysis}

In this section, we will analyse the morphology of the AGN sample identify in this work in order to study differences or similarities between both AGN types. 
We have used the Cassata's morphological catalog \citep{cassata} which provides information of five non-parametric diagnostics of galaxy structure, i.e., asymmetry A, concentration C, Gini coefficient G, second-order moment of the brightest 20\% of galaxy pixels M20 (e.g., \citealt{conselice,abraham,lotz}), for 232022 galaxies up to F814W$=$25, using the Advanced Camera for Surveys (ACS) on board Hubble Space Telescope.

The Gini coefficient is defined as the absolute value of the difference between the integrated cumulative distribution of galaxy intensities and a uniform intensity distribution \citep{abraham} and is a concentration parameter strongly correlated with concentration and surface brightness. However, unlike C, G is independent of the large-scale spatial distribution of the galaxy's light, i.e that it does not require the galaxy to be circularly symmetric \citep{abraham}.
The asymmetry parameter A quantifies the degree to which the light of a galaxy is rotationally symmetric. A is measured by subtracting the galaxy image rotated by 180 degrees from the original image.
M20 describes the second-order moment of the brightest 20\% of the galaxy's flux, defined as the sum of the intensity of each pixel multiplied by the square of the distance from the center of the galaxy for the brightest 20\% of the pixels in a galaxy. This concentration parameter is sensitive to off-axis clumps and so provides information of merger signatures such as multiple nuclei, tidal tails, bars, etc.

We cross-correlated the spatial positions of objects detected in the Laigle et al. catalog with those presented by Cassata et al., using a matching radius of 1 arcsec. From a total of 152 Type 2 AGNs, we find 138 (90\%) matches. Whereas in the case of Type 1 AGNs, we find 100\% matches.
In Figure \ref{giniasym}, left panel, we plot Gini vs. Asymmetry for the sample of Type 1 (filled circles) and Type 2 (open circles) AGNs. We have also include dividing lines between regions of predominantly irregulars, spirals and elliptical morphological types taken from \citep{capak}. The dashed line between irregular and spiral galaxies is defined as log$_{10}$(Asymmetry)$=2.353*$log$_{10}$(Gini)+0.353, while the solid line between spiral and early-type galaxies is defined as log$_{10}$(Asymmetry)$=5.500*$log$_{10}$(Gini)$+0.825$. 
In Figure \ref{giniasym}, right panel, we show the M20 vs. Gini parameters. We have also include dividing lines between regions of mergers, Sb/Sc/Irr and E/S0/Sa galaxy types taken from \citet{lotz08}, according to the following definitions, Mergers: $G >-0.14*$M20$+0.33$, Early(E/S0/Sa): $G <-0.14*$M20$+0.33$, and $G > 0.14*$M20$+0.80$, Late(Sb/Sc/Ir): $G < -0.14*$M20$+0.33$, and $G > 0.14*$M20$+0.80$. 
We note in Figure \ref{giniasym}, left panel, that Type 1 AGN reside in the locus formed by elliptical galaxies or compact objects \citep{capak}. While Type 2 AGNs present a more dispersed distribution. We find that 3.6\% of Type 1 AGN are located in the region occupied by irregular galaxies, 18.8\% are identified with spiral galaxies and 77.6\% with elliptical galaxies. While the majority (80\%) of Type 1 objects are located in the region occupied by elliptical or compact objects. Only 2\% are located in the region of spiral galaxies.
In Figure \ref{giniasym}, right panel, we find that 7.7\% of Type 1 AGNs are located in the region identified by mergers. This result is in agreement with those found by \citet{chang}, who studied a sample of $0.5<z<1.5$ AGNs selected by their mid-IR power-law emission. These authors do not find a high merger rate in an obscured AGN sample. 
We also find that 37\% of the total Type 1 sample are identified with spiral galaxies of Sb/Sc/Irr types, and 55.3\% occupy the region identified with elliptical galaxies. We notice however that the majority of the Type I AGNs are close to the merger region compared to the distribution of Type II AGNs.

 \subsection{Corrected Colours and Masses}

It is known that the presence of an AGN can introduce biases in the measurements of some host galaxy parameters, such as colours, masses, etc. In our case, \citet{laigle} provide colours, stellar masses and star formation rates uncorrected by the presence of nuclear AGN light contamination. 
Nevertheless, \citet{pierce} analysed the effect of AGN light in a sample of X-ray-selected AGN host galaxies at $z\sim1$ finding that integrated optical and UV-optical colours are not significantly affected except in extreme cases ($< 10$ \%) where the AGN is very luminous, unobscured, and/or visible as a point source. 
Also, \citet{hickox09} analyse different host galaxy properties of AGNs selected by its emission in radio, X-rays and IR wavebands. They calculated the correction for AGN contamination in the $u-r$ host galaxy colours finding that the typical correction for nuclear contamination ranges from 0 to 0.3 mag.
Similar results were found by \citet{kauff07} for a sample of low redshift spectroscopically selected AGNs. 
These authors compared UV ($NUV-R$) and optical ($g-r$) colours from the central regions to integrated or total colours. They found that the colours are not strongly affected by light from an AGN, i.e that light from stars in the outer regions of these AGN host galaxies dominates the observed UV-optical colours.
\begin{figure}
 \centering 
\includegraphics[width=0.47\textwidth]{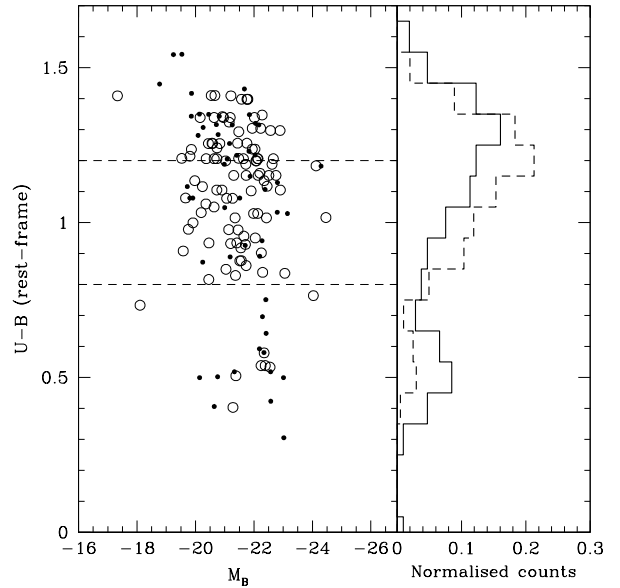}
	\caption{Left panel: Rest-frame $U-B$ vs. $M_B$ colours of Type 1 (filled circles) and Type 2 (open circles) AGNs. Horizontal dashed lines represent the region occupied by ``green valley" galaxies (0.8$\leq U-B \leq $1.2). Right panel: colour distribution of different AGN types. Solid and dashed line histograms represent the colour distribution of Type 1 and 2 AGNs, respectively.}
\label{ub}
\end{figure}

We also carried out a visual inspection of high resolution (0.03"/pixel) HST images (F814W filter) of our sample of AGNs finding that only 12\% of type I AGNs present visible nuclear point source in their optical images \footnote{We have used the Cutouts Service from the NASA/IPAC Infrared Science Archive http://irsa.ipac.caltech.edu/data/COSMOS/index\_cutouts.html}.
Following the previous results, we estimate that the correction for AGN contamination on host galaxy colours is less than $\sim$0.3 mag. and so the results are not affected.

Despite this, in this and the following subsections we will used U-B colours and stellar masses corrected for AGN light contamination.
For this reason, we cross-correlated our AGN samples with the catalog presented in \citet{bongio12}. These authors calculated colours, stellar masses and star-formation rates (SFR) of a sample of X-ray AGNs selected from the $XMM$-COSMOS catalog. These parameters were calculated on a careful study of their spectral energy distributions, which have been parametrized using a two-component (AGN$+$galaxy) model fit.
From our total AGN sample, we find that 101 (70\%) and 50 (87\%) Type 2 and 1 AGNs are in the catalog presented by \citet{bongio12}. 

As mentioned by \citet{bongio12}, there is a problem with SFR determinations for Type 1 AGNs.
These authors find that the accretion disc of Type 1 AGNs contributes significantly to the emission in the UV bands,
which causes a degeneration in the SED fitting method of unobscured objects when considering between the UV emission from the star-formation of the host galaxy and that produced by the central AGN. For this reason we have not performed any analysis with this parameter for the case of AGNs.

Host galaxy colours are important for revealing the nature of AGN host galaxies. It is well-known that the distribution of optical colours in normal galaxies is bimodal, where the predominantly red early-type galaxies occupy a distinct locus in colour from the blue star-forming galaxies \citep{baldry,salim}. It can also be seen an intermediate region called the ``green valley", the region proposed to be the transitional phase in galaxy evolution where galaxies are moving from later-to earlier-types \citep{martin05,martin07}.

In Figure \ref{ub} we plot the rest-frame $U-B$ vs. $M_B$ corrected colours of Type 1 and Type 2 AGNs (left panel).
We have also included the criteria to separate the colours of the ``green valley" galaxies (0.8$\leq U-B \leq $1.2; \citealt{nandra,willmer,mahoro}). In the right panel we show the colour distribution of the different AGN types.
For Type 2 AGNs, we find that 40.5\% are located in the red sequence of bulge-dominated, passively evolving galaxies, 51.5\% are found in the ``green valley" between the red sequence and the blue cloud, and only 8\% are distributed along the blue cloud (generally disk-dominated, star-forming galaxies). While for the sample of Type 1 AGNs, we find 41.5\%, 32\% and 26.5\% located in the red sequence, the ``green valley" and the blue cloud, respectively. As in previous works \citep{hickox09}, we find that our AGN sample lie predominantly in the ``green valley" of the colour–magnitude diagrams.

\subsubsection{Mass vs. redshift}

\begin{figure}
 \centering 
\includegraphics[width=0.47\textwidth]{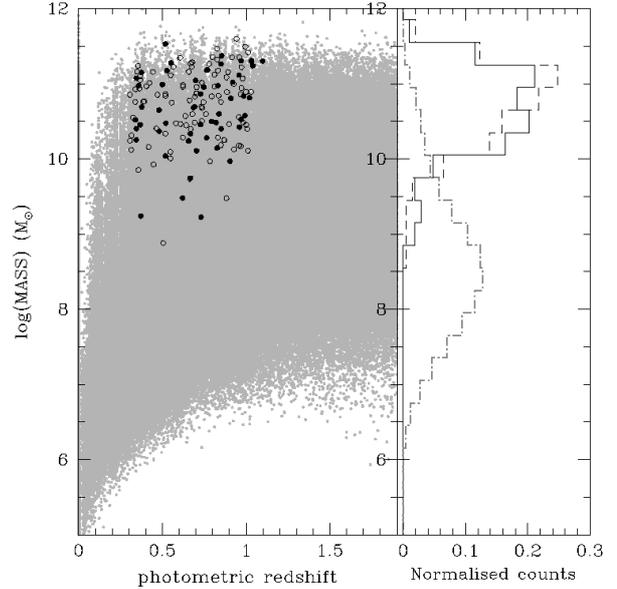}
	\caption{Left panel: Host galaxy stellar masses versus photometric redshift for the parent galaxy sample (gray dots). Filled and open circles represent the corresponding values for Type 1 and 2 AGNs, respectively. Right panel: Stellar mass distribution for Type 1 AGNs (solid line histogram) and Type 2 (dashed line histogram). The dot-dashed line histogram shows the stellar mass distribution for galaxies in the same range of redshifts ($0.3<z<1.1$) as the sample of AGNs.}
\label{mass}
\end{figure}

In Figure \ref{mass} we plot host galaxy stellar masses (MASS\_BEST) versus photometric redshift for the entire galaxy sample (gray dots). We also include the corresponding values for the AGN sample obtained from Bongiorno et al. catalog: open and filled dots represent Type 2 and 1 AGNs, respectively. In the case of AGNs we plot the spectroscopic redshift values. In the left panel we show the corresponding distribution of stellar mass for Type 1 (solid line histogram) and Type 2 (dashed line histogram). We also included the stellar mass distribution for galaxies in the same range of redshifts as the sample of AGNs (dot-dashed line histogram).

As it can be seen, AGNs reside in the most massive host galaxies at any redshift and both AGN types show similar host galaxy stellar mass distributions. For the Type 1 sample we find a mean value $\mu_1=10.6$ and a standard deviation $\sigma_1=0.5$. For the Type 2 sample we find $\mu_2=10.8$ and $\sigma_2=0.3$. For galaxies in the same range of redshifts as the sample of AGNs, we find $\mu_g=8.8$ and a standard deviation $\sigma_g=1$.


\section{Type 1 and 2 AGN environments}

\subsection{Galaxy density}

It is well-known that galaxy environment has a fundamental effect on the properties of galaxies and their evolution \citep{darvish15,muldrew}. ZZ If the postulates of the unified model are true, then it would be expected that the galaxy environments of the different AGN types would be similar.
In this Section we analyze the environment of Type 1 and 2 AGNs.

In Figure \ref{pro} we show the projected radial density of tracer galaxies with $|z_{AGN}-z_{galaxy}|\leq0.2$ and $r<26.5$ (which is the 3$\sigma$ detection limits in the Subaru $r$ band images, \citealt{laigle}) around the obscured and unobscured AGNs with $r_p<500$ kpc. Error bars were estimated using Poissonian errors.
The radial distribution of tracer galaxies in both AGN samples are similar, except that in small scales Type 2 AGNs have more neighbouring galaxies than the sample of Type 1 AGNs.

We have also performed tests using other redshift limits, changing the redshift difference cut to $|z_{AGN}-z_{galaxy}|\leq0.1$ and $0.15$ does not change the results except that the number of tracer galaxies is reduced.

\subsection{Neighbour galaxy properties}

Several studies have shown that galaxy environment correlate with galaxy properties such as morphology \citep{dressler,postman,alpa}, colour \citep{tanaka,balogh}, and SFR \citep{grutz}. 

\begin{figure}
 \centering 
\includegraphics[width=0.47\textwidth]{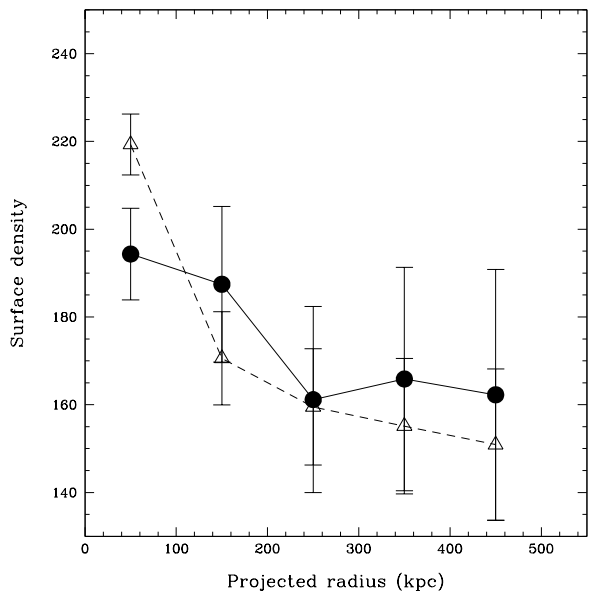}
\caption{Projected radial density of tracer galaxies with $|z_{AGN}-z_{galaxy}|\leq0.2$ and $r<26.5$ around the selected Type 1 (filled circles) and Type 2 (open circles). The error bars represent the standard deviation within each data bin, estimated using Poissonian errors.}
\label{pro}
\end{figure}

\begin{figure*}
\includegraphics[width=55mm]{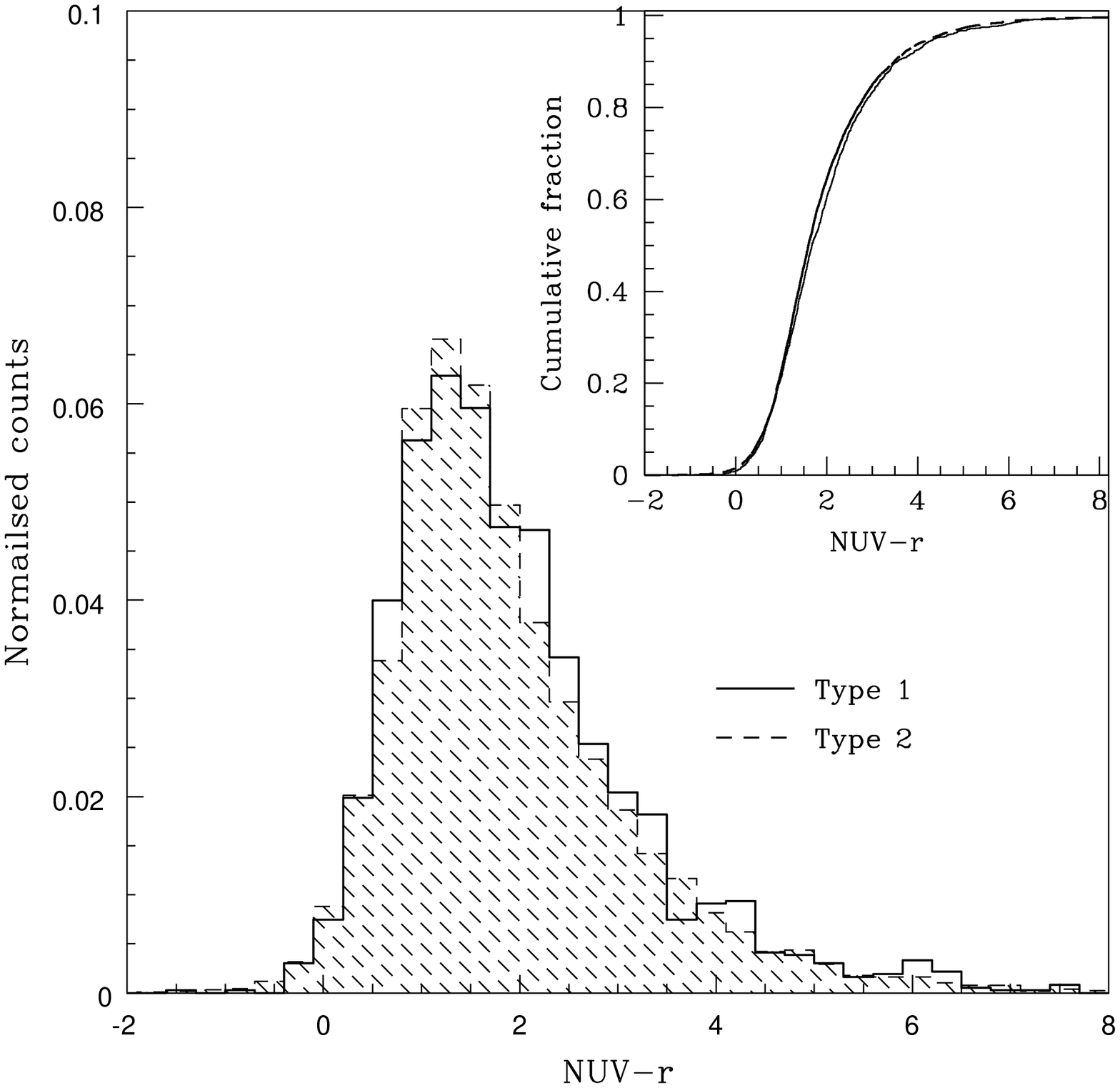}
\includegraphics[width=55mm]{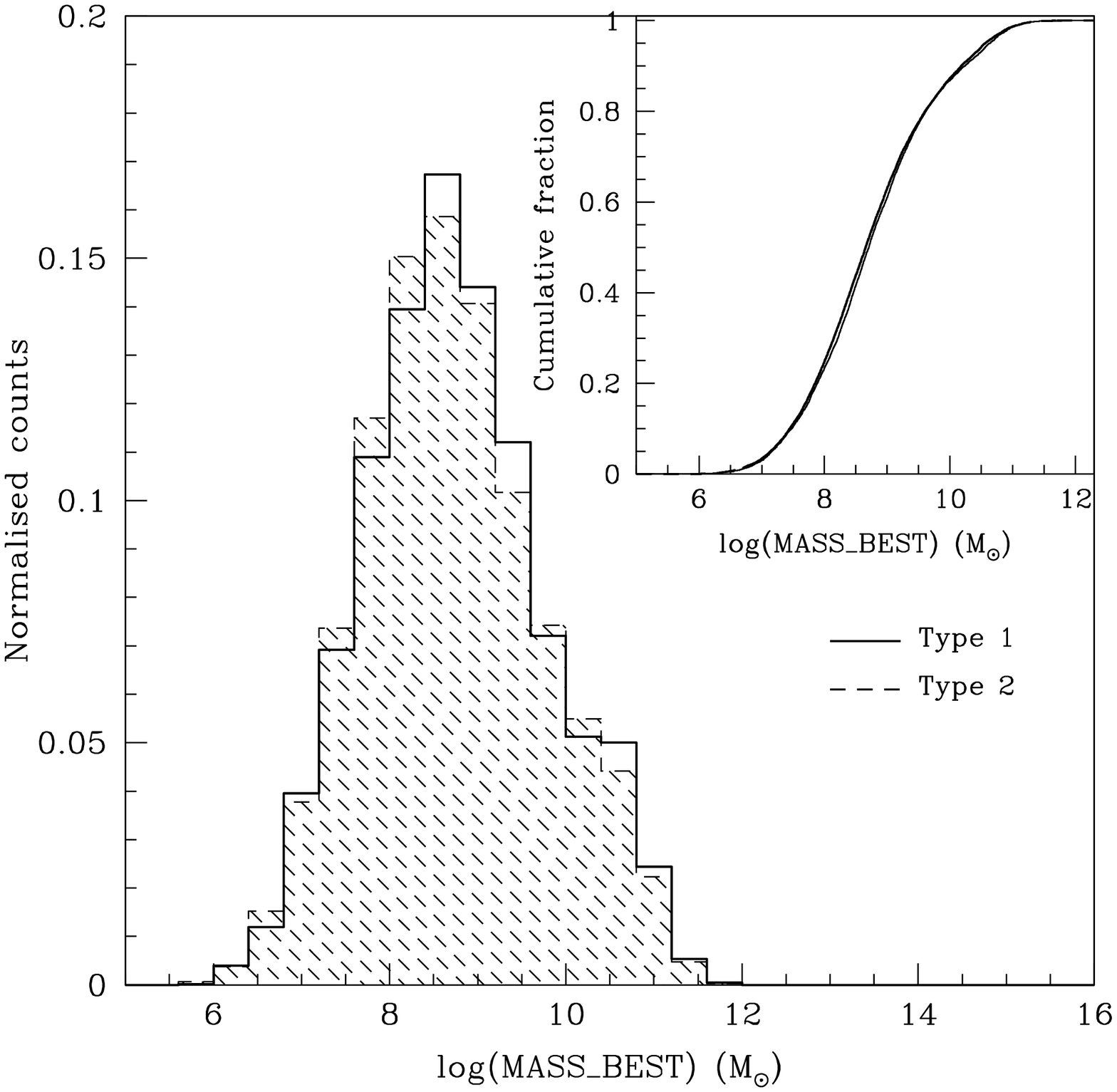}
\includegraphics[width=55mm]{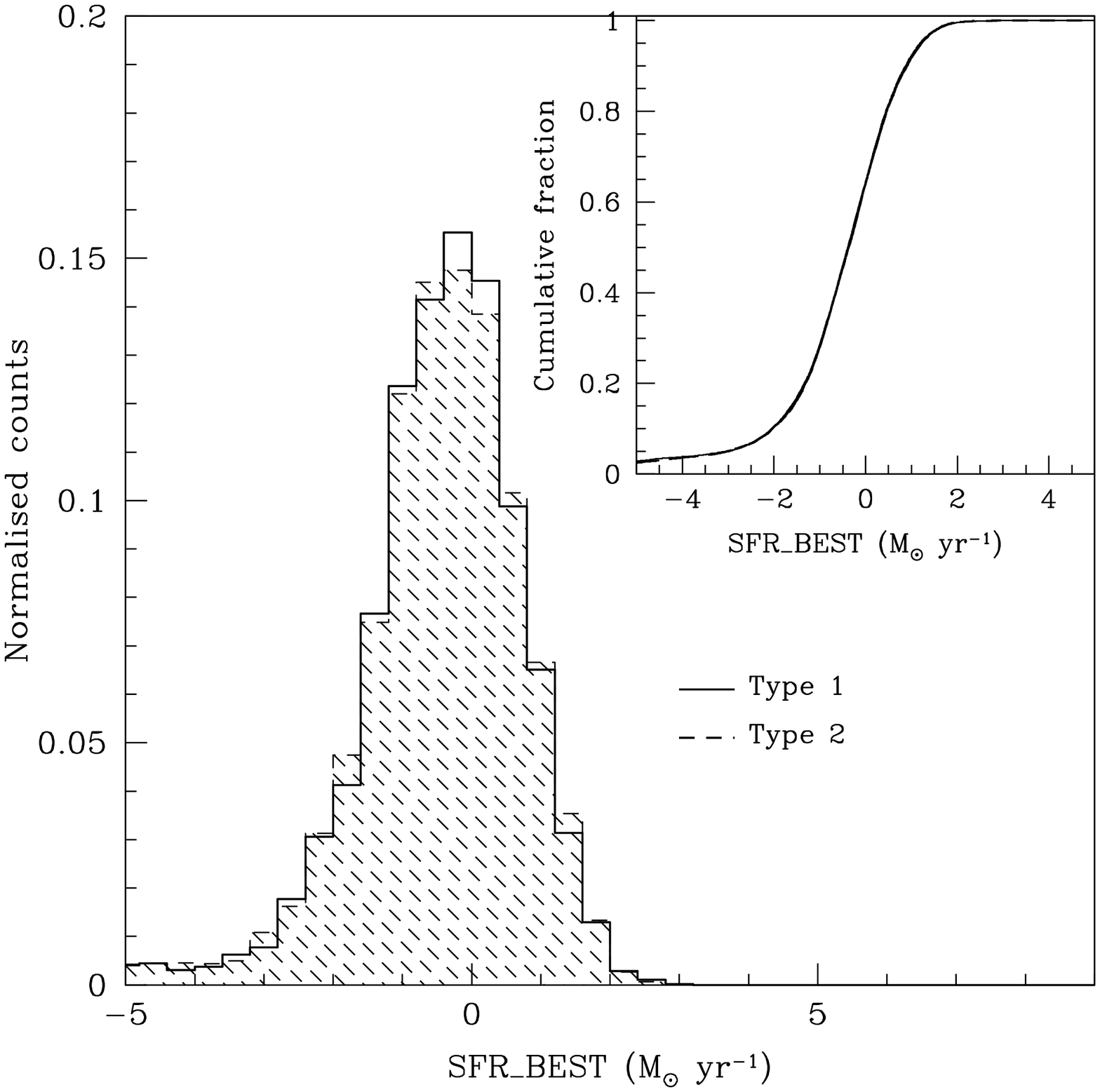}
\caption{NUV-r colour (left panel), stellar masses (middle panel), and SFR (left panel) for tracer galaxies with $|z_{AGN}-z_{galaxy}|\leq0.2$ within 500 kpc from Type 1 (solid line histogram) and Type 2 (shade histogram) AGNs. The inset shows the cumulative fraction distributions.}
\label{proper}
\end{figure*}

\begin{figure*}
\includegraphics[width=55mm]{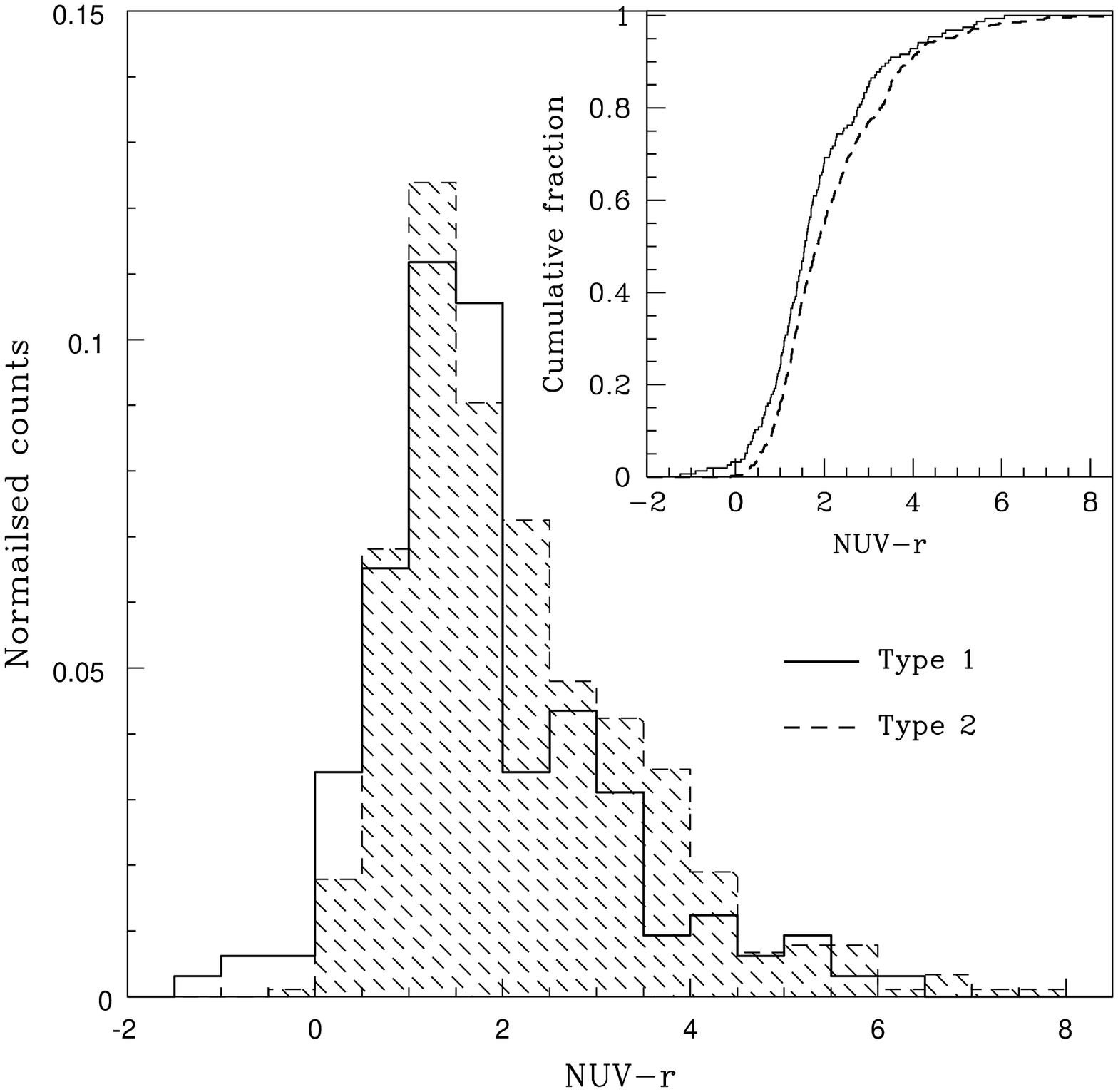}
\includegraphics[width=55mm]{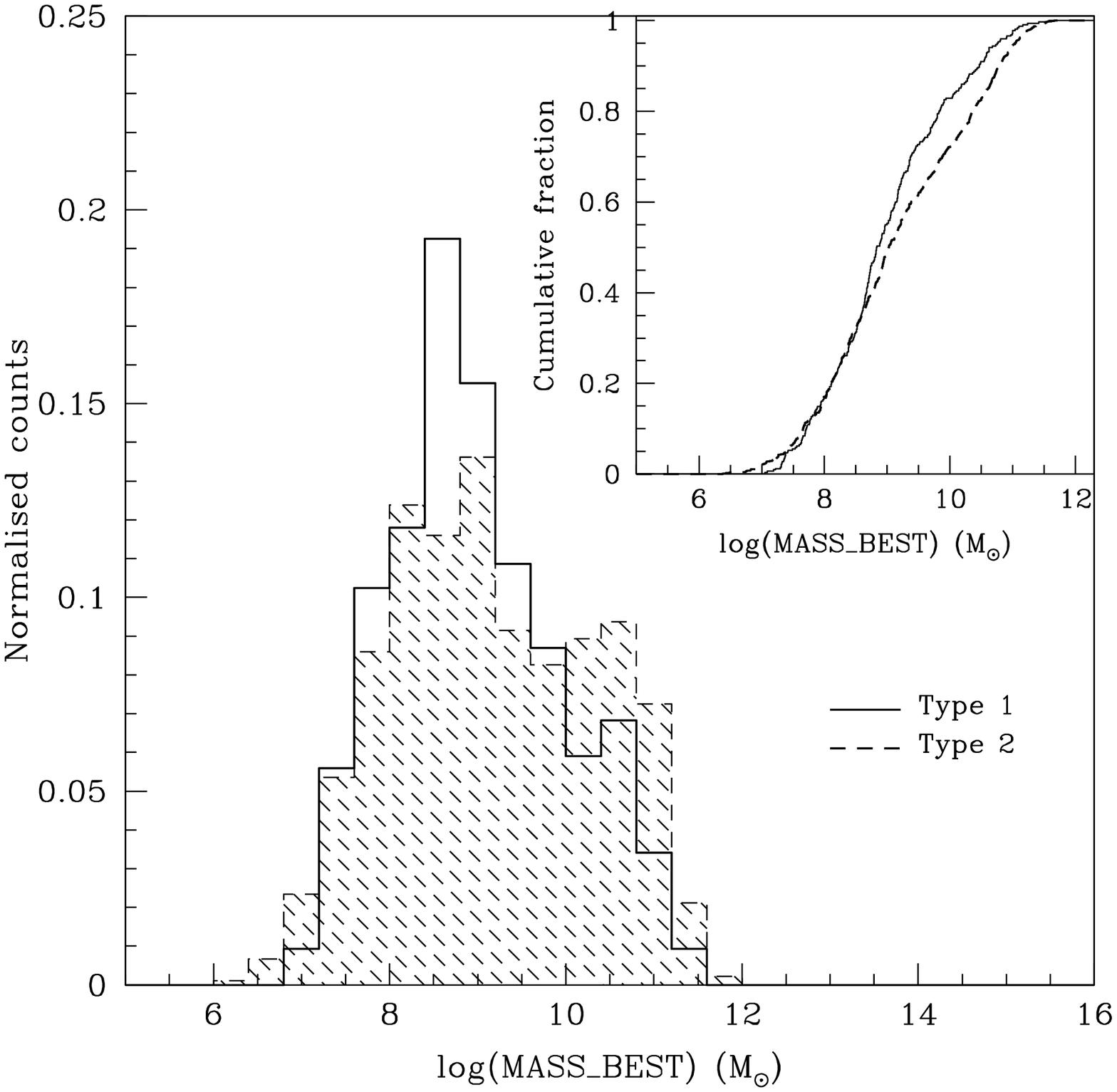}
\includegraphics[width=55mm]{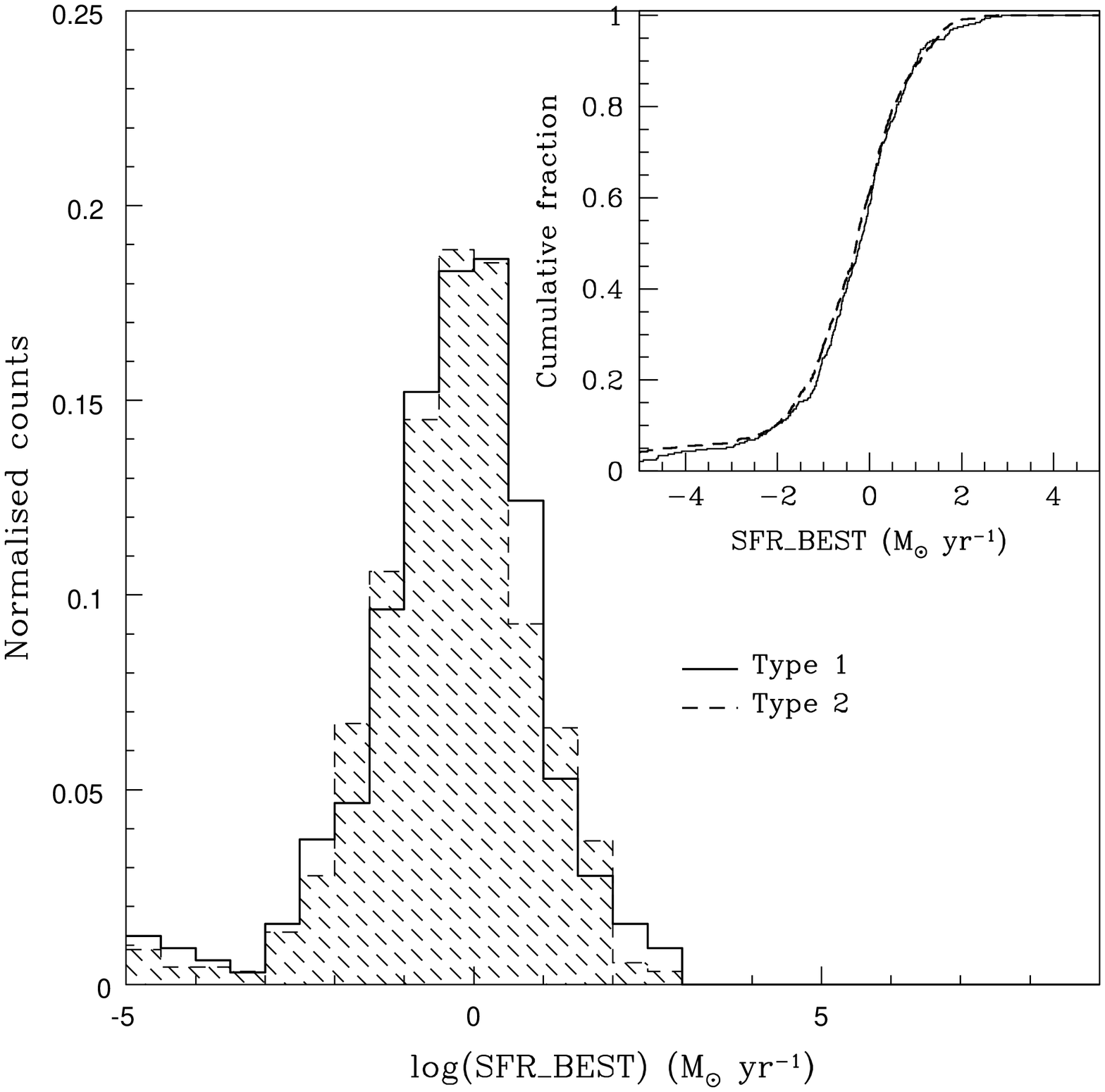}
\caption{Same as Figure \ref{proper} for a sample of tracer galaxies within $100$ kpc from the sample of AGNs}
\label{100}
\end{figure*}

Here we study colour, masses, and SFR of neighbour tracer galaxies located in the field of Type 1 and 2 AGN samples.
In Figure \ref{proper} we show the $NUV-r$ colour (left panel), stellar masses (MASS\_BEST, middle panel), and SFR (left panel) for tracer galaxies with $|z_{AGN}-z_{galaxy}|\leq0.2$ and $r<26.5$ within 500 kpc from Type 1 and Type 2 AGNs. 
As it can be seen, the colours of tracer galaxies at $r_p<500$ kpc show no difference between both AGN samples.
We have included the cumulative fraction distribution for these samples (top right on each figure), in order to make clearer the differences between the different distributions.
Following the results shown in Figure \ref{pro}, we have studied the distribution of the same parameters on smaller scales $r_p<100$ kpc. Contrary to what was found at larger projected radius, differences in the colour and mass distribution of tracer galaxies (see Figure \ref{100}) are clear. Small-scale environments around AGN reveals that neighbour galaxies in the field of Type 2 AGNs tend to have red $NUV-r$ colours compared with the galaxy environment of Type 1 AGNs. This result implies that at larger scales the colour of tracer galaxies around the sample of AGNs are similar, and the differences are only significant at small scales, $r_p<100$ kpc, around both AGN samples.
In Figure \ref{proper}, right panel, we can see that there is no difference between the SFR distribution of tracer galaxies in both AGN samples at larger scales ($r_p< 500$ kpc).


\section{Summary and discussion}

In this paper we study the host galaxies properties and environment of a well designed sample of Type 1 and 2 AGNs at high redshifts ($0.3<z<1.1$) from the COSMOS2015 survey identified on the basis of their spectral, photometric and X-ray emission properties. 

The main results are:

\begin{itemize}

\item Type 1 and 2 AGNs exhibit different colour distribution in the mid-IR and in the optical to mid-IR. Type 1 AGN have blue  $U-[4.5]$ colours in comparison with the Type 2 sample. We observed that the  optical to mid-IR colour distribution of Type 2 objects present a bimodality, indicating two different host galaxies populations. While the $[3.6]-[4.5]$ colour distribution appear inverted, i.e. Type 1 AGNs are bluer comparing to the Type 2 sample. 

\item From the rest-frame $M_{NUV}-M_R$ vs. $M_R-M_J$ plane, we find that different AGNs are located in two particular regions: Type 1 are identified in the regions occupied by star-forming galaxies, whereas 29\% of Type 2s are identified in the locus occupied by quiescent galaxies, while the rest belong to the region populated by star-forming galaxies.

\item According to the analysis carried out in the X-rays, we find that most of the AGNs have X-ray luminosities similar to those found in  Seyfert-like galaxies (with $10^{42}\leq L_X<10^{44}$ \ergs), while only a minority have X-ray luminosities as the observed in QSOs ($L_X > 10^{44}$ \ergs). The Hardness ratio values of Type 1 AGNs are also in agreement with the expected values for unobscured X-ray sources.

	
\item Morphological analysis according to non-parametric measurements (Gini vs. Asymmetry) shows that the majority of Type 1 AGNs are located in the region occupied by elliptical or compact galaxies. While Type 2 AGNs present more scatter, from those occupied by spiral, irregulars and elliptical galaxies.
From the Gini vs. M20 plane, we find that only a small fraction (3.8\%) of the total sample of AGNs are located in the region of mergers. While the rest is located almost equally distributed between the region occupied by early types (E/S0/Sa) and late types (Sb/Sc/Irr). 
Recently, \citet{chang} presented an analysis of IR-selected AGNs in the COSMOS survey. They found also that only half of the most powerful AGNs present evidence of galaxy major mergers. These authors postulate that AGN activity might be triggered by internal mechanisms, such as secular processes, disk instabilities, and compaction in a particular evolutionary stage.

\item { No significant difference is found between the Type 1 and Type 2 AGN host stellar masses, which on average show the same host galaxy mass distribution. We find that both AGN types represent the most massive galaxies at any redshift.}

\item The environment of the different AGN types are similar. The only difference is found at small scales ($r_p<$100 kpc) around the Type 2 sample. Type 2 sources have more neighboring galaxies than Type 1s.

\item We study colour, masses, and star-formation rates (SFR) of neighbour tracer galaxies located in the field of Type 1 and 2 AGN samples. Neighbouring galaxies located within 100 kpc and with $|z_{AGN}-z_{galaxy}|\leq0.2$ from Type 2 AGNs are redder and more massive than the neighbouring galaxies in the field of Type 1 sources. We do not find any differences in the SFR distribution of tracer galaxies in both AGN samples. 

\end{itemize}

These results suggest that the host galaxies of Type 1 and 2 AGNs present different optical, mid-IR, X-ray and morphological properties and also different galaxy environments. These last results are in agreement with those found by \citet{jiang}, previously commented at low redshifts and \citet{koulour}, who found that the fraction of Seyfert 2 galaxies with a close neighbour (within a projected distance of 100 kpc) is significantly higher than that of their control and Seyfert 1 galaxy samples. Similar results were found by \citet{dultzin} who observed a significant excess of large companions galaxies within a radius of 100 kpc.
Other studies also find differences in the environment of distinct AGNs at high redshifts. Although the selection criteria vary from one job to another, several works found that Type 2 (obscured) and Type 1 (unobscured) AGNs reside in different galaxy environment. These differences are also observed in the auto cross-correlation function calculated for a sample of obscured and unobscured AGNs selected from mid-IR colour cuts at high redshift (\citealt{hickox11,donoso,dipompeo}).

We think that the AGN obscuration observed in Type 2s is possibly due to galaxy interactions or mergers. 
Although the results found in the morphological analysis show that only a small fraction of the AGNs are mergers, we consider the possibility that after galaxy mergers, there are still large amounts of dust that obscure the central engine. 
This proposal is in agreement with the results obtained by \citet{gould12}, who concluded that the dust located in the host galaxy, in a sample of local Compton-thick AGNs, is responsible for the dominant contribution to the observed mid-IR dust extinction and not necessarily a compact obscuring torus surrounding the central engine.

We find different AGN host galaxy properties, environments and neighboring galaxies in a sample of Type 1 and 2 AGNs.
These results are more related to evolutionary sequences than to the Unified model schemes. The obscuration due to the presence of gas and dust can be distributed in larger galactic-scales and can be due possibly to an evolutionary process produced by galaxy interactions or mergers.







\section*{Acknowledgments}
We are grateful to the anonymous Referee for his/her careful reading of the manuscript and a number of comments, which improved the the quality of this manuscript.
Based on data products from observations made with ESO Telescopes at the La
Silla Paranal Observatory under ESO programme ID 179.A-2005 and on
data products produced by TERAPIX and the Cambridge Astronomy Survey
Unit on behalf of the UltraVISTA consortium.
Based on data obtained with the European Southern Observatory Very Large Telescope, Paranal, Chile, under Large Programs 175.A-0839 (zCOSMOS), 179.A-2005 (UltraVista) and 185.A-0791 (VUDS).
This work was partially supported by the Consejo Nacional de Investigaciones Cient\'{\i}ficas y T\'ecnicas (CONICET) and the Secretar\'ia de Ciencia y Tecnolog\'ia de la Universidad de C\'ordoba (SeCyT).


{}

\end{document}